\documentclass[a4paper,nofootinbib,
preprint,
showpacs
preprint
]{revtex4}
\usepackage{graphicx}
\usepackage{axodraw}
\def\eq#1{{eq.~(\ref{#1})}}
\def\eqs#1#2{{eqs.~(\ref{#1})--(\ref{#2})}}
\def\sec#1{{sec.~(\ref{#1})}}
\def\tab#1{{Tab.~(\ref{#1})}}
\def\fig#1{{Fig.~(\ref{#1})}}

\def\vev#1{\left\langle #1\right\rangle}

\def\Re{\mbox{Re}\,}
\def\Tr{\mbox{Tr}\,}

\def\hbar{\hspace{0pt}\raisebox{1pt}{$-$} \hspace{-7pt} h}

\def\5{\overline 5}

\newcommand{\be}{\begin{equation}}
\newcommand{\ee}{\end{equation}}
\newcommand{\bea}{\begin{eqnarray}}
\newcommand{\eea}{\end{eqnarray}}
\newcommand{\nn}{\nonumber}

\begin{document}
\title[Phenomenology of the little flavons model]{Phenomenology of the little flavon model
}
\date{\today}
\author{Federica Bazzocchi}
\affiliation{Scuola Internazionale Superiore di Studi Avanzati\\
and INFN, Sezione di Trieste \\
via Beirut 4, I-34014 Trieste, Italy}
\begin{abstract}
The phenomenology of the little-flavon model  is discussed. Flavor changing neutral current and lepto-quark compositeness set the most stringent bounds to the lowest possible value for the scale at which the flavons arise as pseudo-Goldstone bosons.
\end{abstract}
\maketitle
%
\vskip1.5em
\section{Motivations}
The little-flavon model \cite{noi} explains fermion masses and mixing matrices  in a little-Higgs inspired scenario \cite{littlehiggs}. In the model, an $SU(2)_F\times U(1)_F$
gauge flavor symmetry is spontaneously and completely broken by the vacuum of the dynamically induced potential for  two scalar doublets, the little-flavons, which are pseudo-Goldstone bosons
remaining after the spontaneous breaking---at a scale $\Lambda = 4\pi f$---of an approximate $SU(6)$ global symmetry. The  $SU(2)_F\times U(1)_F$ flavor symmetry is the diagonal combinations of a $[SU(2)\times U(1)]^2$ gauge symmetry surviving
 the spontaneous breaking of $SU(6)$.

The model reproduces successfully fermion masses and mixing angles \cite{noi}.
Ratios between the values of the vacuum expectation values of the flavons and $f$ give
 rise to the  textures in the fermion mass matrices. For this reason fermion masses and
 mixing angles predicted by the model are independent of the scale $f$. On the other hand,
  the masses of the flavor bosons---both vector and scalar---arising in the breaking of the
  gauge symmetries depend on $f$. Since these masses enter in the processes mediated by the new particles, a computation of their effects allows us to constrain  possible values of $f$.

In this letter, several processes are considered together with their experimental bounds. We present the relevant processes that occur at tree level starting from those that give the less stringent bounds
on $f$ to that that give the most stringent. The latter bound comes from flavor changing
 neutral current in the $K^0$-$\bar{K}^0$ for which we have $\Lambda\geq 5\times 10^4$ TeV.
 We also give an example of a one-loop process that gives a limit on $f$ as stringent
  as that coming from $K^0$-$\bar{K}^0$. Finally we show how the electroweak Higgs mass scale
  ($\simeq 200$ GeV) can be kept stable notwithstanding the much higher scale of flavor physics.

\section{Interactions}
\label{inte}
The effective lagrangian at the scale $\Lambda$ is
given by \be \mathcal{L}= \mathcal{L}^{\Sigma}_{kin}
+\mathcal{L}^{f}_{kin}+ \mathcal{L}^{g}_{kin} + \mathcal{L}_Y \,,
\label{L}
\ee
where $\mathcal{L}^{\Sigma,f,g}_{kin}$  includes the
kinetic terms for the pseudo-Goldstone bosons, the fermions and
the gauge bosons respectively and $\mathcal{L}_Y $  the Yukawa
couplings. Explicitly we have \bea \mathcal{L}^{\Sigma}_{kin}&=&
- \frac{f^2}{4} \Tr
\left( D^\mu \Sigma \right) \left( D_\mu \Sigma \right)^\ast\,, \nn\\
\mathcal{L}^{f}_{kin} & =& \bar{f}_{L,R}\gamma_{\mu}(\partial^\mu+
i g_1 A^{\mu}_{1\,a} T^{a}+ i g'_{1}B^\mu_{1} )f_{L,R} \,\,,
\label{int}
\eea
with
\be
 D^\mu \Sigma = \partial^\mu + i g_i A^\mu_{i\,a} \left( Q_i^a \Sigma + \Sigma Q_i^{aT} \right)+ i g'_i B^\mu_{i} \left( Y_i \Sigma + \Sigma Y_i^T \right)\,,
\ee
where $\Sigma= \exp(i\Pi/f)\Sigma_{0}$ is the non linear representations of the pseudo-Goldstone bosons, $A^\mu_{i\,a},\,B^\mu_{i},\,(i=1,2;\, a=1,2,3$), the gauge bosons of the two copies of $SU(2)\times U(1)$ and $g_{i},\,g'_{i}$ their couplings. $\mathcal{L}_Y$ is rather cumbersome and we report its expression in the Appendix.

After the spontaneous breaking of global $SU(6)$ we are left with
four massive gauge bosons, ${A'}^{\mu}_a\, (a=1,2,3)$ and ${B'}^\mu$ of masses
\bea m^2_{A'}= \frac{(g_1^2 +g_2^2)}{2}f^2 &\mbox{and}& m^2_{B'}=
\frac{2}{5}({g'}_1^{2} + {g'}_2^{2})f^2 \,,
\eea
and four massless gauge bosons, $A^{\mu}_a \,(a=1,2,3)$ and $B^{\mu}$.

After the $SU(2)_F\times U(1)_F$ symmetry is broken we are left with one complex massive gauge boson, $F_3^{\mu}$, $2$ real  massive gauge bosons, $F^{\mu}_{1,2}$,  $2$ real, $\varphi_{1,2}$, and one
complex, $\varphi_{3}$, massive scalars, which are the flavon scalars. Their masses
are given by
\bea
m^2_{F_{3}}&=& \frac{1}{2}g^2(\epsilon_1^2 +\epsilon_2^2)f^2\, ,\nn \\
 m^2_{F_{1,2}}&=& \frac{1}{2}(g^2+{g'}^{2})\epsilon_{1,2}^2 f^2\,, \nn\\
m^2_{\varphi_{1,2}}&=& \big[\big(\lambda_1 \epsilon_1^2 + \lambda_2 \epsilon_2^2 \pm \nn \\
&&\sqrt{ (\lambda_1 \epsilon_1^2 + \lambda_2 \epsilon_2^2)^2 -( 4
\lambda_1\lambda_2 -\lambda_3^2) \epsilon_1^2
\epsilon_2^2}\big)f^2\big]\,,\nn \\
m^2_{\varphi_{3}}&=& \frac{1}{2}\lambda_{4}(\epsilon_1^2 +\epsilon_2^2)f^2\, ,
\eea
where $g^2= g_1^{2} g_2^{2}/(g_1^{2}+
g_2^{2})$ , ${g'}^{2}= {g'}_1^{2} {g'}_2^{2}/({g'}_1^{2}+ {g'}_2^{2})$ are
the effective gauge couplings, $\lambda_{4}\simeq \mathcal{O}(1)$ and
$\lambda_{1,2,3}\simeq \mathcal{O}(10^{-2})$ are the parameters of the
potential as discussed in \cite{noi} and $\epsilon_{1,2}$
the ratios of the vacuum expectation values of $\phi_1$ and $\phi_2$ and
the scale $f$. The numerical analysis in \cite{noi}
indicates that  $\epsilon_1\epsilon_2\simeq 0.2$ if we want to fit the fermion masses and mixing angles. The gauge bosons
$F^{\mu}_{1,2}$ come from the mixing between $A^{\mu}_3$ and
$B^{\mu}$, while $F^{\mu}_{3}$ from the mixing between $A^{\mu}_1$ and
$A^{\mu}_2$.

From the kinetic term in \eq{int} we have the following
interactions between the gauge bosons and the fermions \be
y^{f}_{F_{L,R}}\Bigg(\frac{\sqrt{g^2 +
{g'}^{2}}}{\sqrt{2}}\Bigg)\big(\overline{f}_{L,R}\gamma_{\mu}f_{L,R}\big)\big(F^{\mu}_{1}+F^{\mu}_{2}\big)\,,
\label{sing} \ee if $f_{L,R}$ is a singlet of $SU(2)_{F}$ with
flavor hypercharge  $y^{f}_{F_{L,R}}$, and \be
\frac{g}{\sqrt{2}}\Big[\big(\overline{\psi^1}_{L,R}\gamma_{\mu}\psi^2_{L,R}\big)F^{\dag\mu}_{3}
+
  h.c.\Big]+ \frac{\sqrt{g^2 +
{g'}^{2}}}{\sqrt{2}}\Big[\big(\overline{\psi^1}_{L,R}\gamma_{\mu}\psi^1_{L,R}\big)F^{\mu}_{2}
+
\big(\overline{\psi^2}_{L,R}\gamma_{\mu}\psi^2_{L,R}\big)F^{\mu}_{1}
\Big] \, , \label{dop} \ee
 if $\psi_{L,R}$ is a doublet of $SU(2)_{F}$ of flavor
hypercharge $1/2$ with components $\psi^1_{L,R}$  and $\psi^2_{L,R}$.

The interactions in \eqs{sing}{dop} have been written in the flavor current
basis for the fermions $f_{L,R}$ and $\psi_{L,R}$. In the next
sections we will indicate as $e^{i=1,2,3}_{L,R}$ the charged lepton
flavor current eigenstates, $e^{\alpha=1,2,3}_{L,R}=e_{L,R},\mu_{L,R},\tau_{L,R}$ the charged
lepton mass eigenstates, $L^e_{i,\alpha},R^{e}_{i,\alpha}$
the unitary matrices that diagonalize the non diagonal mass matrix $M^{RL}_e$ through the bi-unitary transformation
\be
{R^{e}}^\dag \, M^{RL}_e \,L^e = M^{RL^{diag}}_e \,.
\ee
 The same conventions  will be used for
the quarks, where we have $u^i_{L,R}$, $d^i_{L,R}$, $u^{\alpha=1,2,3}_{L,R}=
u_{L,R},c_{L,R},t_{L,R}$ and $d^{\alpha=1,2,3}_{L,R}=
d_{L,R},b_{L,R},s_{L,R}$,
$L^{u,d}_{i,\alpha},R^{u,d}_{i,\alpha}$ and $M^{RL}_u\,$ and $M^{RL}_d$. For completeness all the non diagonal mass matrices are reported in the Appendix.

In the model all the Standard Model quarks are $SU(2)_F$ singlets
charged under $U(1)_F$. Standard Model leptons belonging to the
first family, $l_{e\,L}$ and $e_R$, are $SU(2)_F$ singlets as
well, while those of the second and third family, $l_{\mu ,\tau
\,L}$ and $(\mu,\tau)_R$, are members of a doublet in flavor space
(see \tab{charge} in the Appendix). A consequence of this choice
is that all lepton mass eigenstates interact with all the gauge
bosons after $SU(2)_F\times U(1)_F$ is completely broken. For this
reason it is useful to write down the interactions between flavor
gauge bosons and charged lepton mass eigenstates. The general
interaction is given by \be
y^{e\alpha\,\beta}_{m_{L,R}}\Bigg(\frac{\sqrt{g^2 +
{g'}^{2}}}{\sqrt{2}}\Bigg)\big(\overline{e}^\alpha_{L,R}\gamma_{\mu}e^\beta_{L,R}\big)F^{\mu}_{m}\,,
\ee where $m=1,2,3$ and $y^{\alpha\,\beta}_{m_{L,R}}$ are given by
\bea y^{e\alpha\,\beta}_{1_{U}} &= &
U^{e*}_{1\,\alpha}U^e_{1\,\beta}y_U^{e1} +
U^{e*}_{3\,\alpha}U^e_{3\,\beta} \,,
\nn\\
y^{e\alpha\,\beta}_{2_{U}} &= &
U^{e*}_{1\,\alpha}U^e_{1\,\beta}y_U^{e1} +
U^{e*}_{2\,\alpha}U^e_{2\,\beta} \,,\nn\\
y^{e\alpha\,\beta}_{3_{U}} &= & \sqrt{\frac{g^2}{g^2 + {g'}^2 }}
U^{e*}_{2\,\alpha}U^e_{3\,\beta}\,, \label{yn} \eea with
$U^e=L^e,\,R^e$ and $y_U^{e1}$ the first
family charged leptons flavor hypercharges (see \tab{charge} in
the Appendix).

For completeness we report also the interaction between flavor
gauge bosons and quark mass eigenstates. The interaction is given by \be
y^{q\alpha\,\beta}_{L,R}\Bigg(\frac{\sqrt{g^2 +
{g'}^{2}}}{\sqrt{2}}\Bigg)\big(\overline{q}^\alpha_{L,R}\gamma_{\mu}q^\beta_{L,R}\big)\big(F^{\mu}_{1}
+ F^\mu_2 \big)\,, \ee where  $y^{q\alpha\,\beta}_{L,R}$ are given
by \be y^{q\alpha\,\beta}_{U} =
\sum_{i=1,2,3}\,U^{q*}_{i\,\alpha}U^q_{i\,\beta}y_U^{q\,i},
\label{ynq} \ee with $U^q=L^q,\,R^q$ and
$y_U^{q\,i}$ the quarks flavor hypercharges (see \tab{charge} in the
Appendix).

\section{Processes mediated by the flavons}
\label{fl}

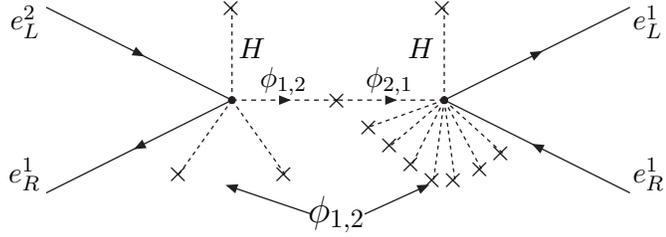
\begin{figure}
\begin{center}
\begin{picture}(300,100)(0,0)
\Vertex(210,50){1.5}
\Vertex(130,50){1.5}
\Text(58,16)[r b]{$e_R^1$}
\ArrowLine(130,50)(60,15)
\Text(58,86)[r
t]{$e_L^2$}
\ArrowLine(60,85)(130,50)
\Text(150,52)[
b]{$\phi_{1,2}$} 
\Text(170,52)[ b]{$$}
\Text(190,52)[
b]{$\phi_{2,1}$}
\DashArrowLine(130,50)(170,50){2}
\DashArrowLine(170,50)(210,50){2}
\Text(170,50)[c]{$\times$}
\Text(282,86)[l
t]{$e_L^1$}
\ArrowLine(210,50)(280,85)
\Text(282,16)[l
b]{$e_R^1$}
\ArrowLine(280,15)(210,50)
\DashLine(130,50)(130,85){1.5}
\Text(130,85)[c]{$\times$}
\Text(133,70)[l]{$H$}
\DashLine(210,50)(210,85){1.5}
\Text(210,85)[c]{$\times$}
\Text(207,70)[r]{$H$}
\DashLine(130,50)(110,22){1.5}
\DashLine(130,50)(150,22){1.5}
\DashLine(210,50)(198,26){1.5}
\DashLine(210,50)(190,33){1.5}
\DashLine(210,50)(182,40){1.5}
\DashLine(210,50)(206,20){1.5}
\DashLine(210,50)(214,20){1.5}
\DashLine(210,50)(224,24){1.5}
\DashLine(210,50)(232,30){1.5}
\Text(198,26)[c]{$\times$}\Text(190,33)[c]{$\times$}
\Text(182,40)[c]{$\times$} \Text(206,20)[c]{$\times$}
\Text(214,20)[c]{$\times$} \Text(224,24)[c]{$\times$}
\Text(232,30)[c]{$\times$} \Text(110,22)[c]{$\times$}
\Text(150,22)[c]{$\times$} \Text(170,0)[b]{\large$\phi_{1,2}$}
\LongArrow(160,7)(129,17.5)
\LongArrow(180,7)(203,17.5)\end{picture}\end{center}\caption{Flavon
mediated contribution to the decay
$\mu^-\,\rightarrow\,e^+\,e^+\,e^-$. The fields $\phi_{1,2}$ are the $SU(2)_F$ doublets.}
\label{bounds}
\end{figure}

All interactions between fermions and flavons come from the Yukawa lagrangian $\mathcal{L}_Y$ in \eq{L} (see the Appendix for the full expression). These are the terms that give origin to the fermion mass matrices. After the breaking of $SU(2)_F\times U(1)_F$ it gives also the interactions we are interested here. Notice that in the following, for simplicity, we will indicate as \textit{flavons} both the $SU(2)_F$ doublets, $\phi_{1,2}$, and the massive scalars, $\varphi_{1,2,3}$, arising after the breaking of the $SU(2)_F\times U(1)_F$ symmetry. Processes mediated by the flavons can occur at tree level and at one or more loops. Tree level processes concern direct interactions between fermions and only
  one flavon, for this reason couplings of this kind  will follow the fermion mass matrices and  all the flavor changing processes mediated by the flavons will be very suppressed since they will result be proportional to power of the ratio between the light fermion masses  and the scale $f$.

In trying to constraint the flavon masses, let us first consider
the lepton flavor violation  (LFV) process  $\mu\rightarrow 3e$.
The limit on the branching ratio $\Gamma_{\mu^-\rightarrow e^{+}
e^{+}e^{-}}$ is given as a function of the total branching ratio
$\Gamma_{\mu\rightarrow \mbox{\small all}}$ \cite{muegamma} \be
\frac{\Gamma_{\mu\rightarrow 3e}}{\Gamma_{\mu\rightarrow
\mbox{\small all}}}< 10^{-12}, \label{exp3e} \ee with \be
\Gamma_{\mu\rightarrow \mbox{\small all}}= \frac{ m_\mu^5
G_F^2}{192\pi^3}. \label{gammaall}\ee
 In the model
we have tree-level LFV processes mediated by the flavons which
give rise to effective operators. They can be parametrized as \bea
&&\frac{1}{\tilde{\Lambda}^2}\Big\{\eta_{LL}\big(\bar{e}(1-\gamma_5)
\mu\,\bar{e}(1-\gamma_5) e\big) +
\eta_{RR}\big(\bar{e}(1+\gamma_5) \mu\,\bar{e}(1+\gamma_5)
e\big)+ \nn\\
&&\eta_{LR}\big(\bar{e}(1-\gamma_5) \mu\,\bar{e}(1+\gamma_5)
e\big)+\eta_{RL}\big(\bar{e}(1+\gamma_5) \mu\,\bar{e}(1-\gamma_5)
e\big)\Big\}\,, \label{parmu}
 \eea
 where $\tilde{\Lambda}$ is an effective scale given by
\be \frac{1}{\tilde{\Lambda}^2}= \frac{1}{4 (4\lambda_1\lambda_2 -\lambda_3^2)(\epsilon_1\epsilon_2)^2 f^2}
\label{flavonsp}
\ee
and
\be
  \eta_{LL} = \Bigg(\frac{R^{e*}_{i1}\, M^{RL}_{e_{ij}} \,L^e_{j2}}{f}\Bigg)\Bigg(\frac{R^*_{l1}\, M^{RL}_{e_{lk}} \,L_{k2}}{f}\Bigg)F_{ijlk}(\lambda_1,\lambda_2,\lambda_3,\epsilon_1,\epsilon_2) \,,
  \label{flavonss}
  \ee
with similar expressions for $\eta_{RR}$, $\eta_{LR}$, $\eta_{RL}$. 
Notice that the effective scale  $\tilde{\Lambda}$ is obtained summing on the exchanges of the two lighter massive flavons, $\varphi_{1,2}$, that are the only ones which give rise to tree level processes. $M_e^{RL}$ in \eq{flavonss} is the non diagonal charged lepton mass matrix (see the Appendix), $L^e$ and $R^e$ are defined in \sec{inte}, $F_{ijlk}$ is a function of the potential parameters $\lambda_{i=1,2,3}$ discussed in \cite{noi} and of $\epsilon_{1,2}$ that depends on the processes multiplicities in the current basis.

From \eqs{flavonsp}{flavonss} we can readily compute
$\Gamma_{\mu\rightarrow e^{+} e^{+}e^{-}}$ that is given by \be
\Gamma^{flavoni}_{\mu\rightarrow 3 e }= \frac{
m_\mu^5}{6(16\pi)^3}\Bigg(\frac{1}{(4\lambda_1\lambda_2-\lambda_3^2)(\epsilon_1\epsilon_2)^2\,f^2}\Bigg)^2\big(|\eta_{LL}|^2 + |\eta_{RR}|^2 + |\eta_{LR}|^2 +
|\eta_{RL}|^2\big)\,. \label{gammaflavmueee}\ee By imposing the
experimental bound in \eq{exp3e} and using \eq{gammaall} we find
\be f>200\, \mbox{GeV}. \ee Such a rather weak bound is justified
by the strong suppression of this process. This is best understood
by going back to the current eigenstates. In this basis we have
nine processes that sum to give $\mu\rightarrow 3 e$ in the mass
eigenstates. For simplicity we consider only one of them. The
interaction terms that give rise to the tree level process are \be
\lambda_{2e}\overline{e}_R^1\, \big(
h^{0*}e_L^2\big)\Big(\frac{\phi_2^\dag
\phi_1}{f^2}\Big)\frac{\phi_1^0}{f} +
\lambda_{1e}\overline{e}_R^1\, \big(h^{0*}
e_L^1\big)\Big(\frac{\phi_2^\dag
\phi_1}{f^2}\Big)^4+\lambda_1(\phi_1^\dag \phi_1)^2 +
\lambda_2(\phi_2^\dag \phi_2)^2 + \lambda_3(\phi_1^\dag
\phi_1)(\phi_2^\dag \phi_2) \,.\ee After the flavor and the
electroweak spontaneous breaking the tree level effective coupling
is (see \fig{bounds}) \be
\frac{\lambda_{2e}\lambda_{1e}}{4(4\lambda_1\lambda_2-
\lambda_3^2) (\epsilon_1\epsilon_2)^2
f^2}\epsilon_1(\epsilon_{1}\epsilon_2)^5
\frac{\vev{h^{0*}}^2}{f^2}(\overline{e}(1-\gamma_5)\mu)(\overline{e}(1-\gamma_5)e)(16\lambda_1
+8\lambda_2 +12\lambda_3) \,, \label{basec} \ee where
$(16\lambda_1 +8\lambda_2 +12\lambda_3)$ is the function indicated
as $F_{1211}$ in \eq{flavonss}. Processes mediated by the flavons
are suppressed by powers of $\epsilon_{1,2}$ and by the ratio
between the electroweak breaking scale and the flavor one.

Let us consider also a flavor changing neutral current (FCNC)
process mediated by the flavons in the quark sector. FCNC
processes in the quark sector with $\Delta F=2$ are responsible of
meson-antimeson oscillations. Since meson mass eigenstates are a
combination of  mesons in the current basis, the splitting of the
masses of the mass eigenstates is related to the possible FCNC
processes. This statement is general and can be applied to
$K^0$-$\bar{K}^0$ as well as $B^0$-$\bar{B}^0$ system.
Nevertheless, the best experimental data are related to the
splitting of Kaon masses \cite{deltamls} \be \Delta m_{LS} =
(3.46\pm0.01)\,\times\, 10^{-12}\, \mbox{MeV}, \label{mslexp} \ee
and therefore we  will consider only the processes with  $\Delta
S=2$.
Given an effective interaction $\mathcal{V}= C\mathcal{O}_{\Delta
S=2}$, where $C$ is a numerical coefficient and
$\mathcal{O}_{\Delta S=2}$ the effective operator involving the
quarks $d$ and $s$, we have that \be \Delta m_{LS} = 2 C
\,\frac{\Re{\langle K^0|\mathcal{O}_{\Delta
S=2}|\bar{K}^0\rangle}}{2m_K} \,.\label{msl} \ee

In order to estimate the contribution of the flavons to $\Delta
m_{LS}$ we have to consider all the possible effective operators
with $\Delta S = 2$. We have three main operators that we
parametrize as follows \bea &&
-\frac{1}{\tilde{\Lambda}^2}\Big[\rho_{1}\Big(\bar{d}(1+\gamma_5)s
\,\bar{d}(1-\gamma_5)s\Big)+ \rho_{2}\Big(\bar{d}(1-\gamma_5)s
\,\bar{d}(1-\gamma_5)s\Big) \nn \\
&& +\rho_{3}\Big(\bar{d}(1+\gamma_5)s
\,\bar{d}(1+\gamma_5)s\Big)\Big] \,, \label{kkopfl} \eea where \be
\frac{1}{\tilde{\Lambda}^2}= \frac{2(\lambda_1
+\lambda_2+\lambda_3)}{4 (4\lambda_1\lambda_2
-\lambda_3^2)(\epsilon_1\epsilon_2)^2 f^2}. \label{flavonspkk} \ee
The coefficients $\rho_i$ are given by 
\bea 
\rho_1 &=& \Bigg(\frac{\Sigma_{ji} (R^{d*}_{j1}M^{RL}_{d_{ji}} N_{ji}
L^{d}_{i2})}{f}\Bigg)
\Bigg(\frac{\Sigma_{lk} (R^{d*}_{l2}M^{RL}_{d_{lk}} N_{lk} L^{d}_{k1})^*}{f}\Bigg)\,,\nn \\
\rho_{2}&=& \Bigg(\frac{\Sigma_{ji} (R^{d*}_{j1}M^{RL}_{d_{ji}} N_{ji} L^{d}_{i2})}{f}\Bigg)^2\,,\nn \\
\rho_3 &=& \Bigg(\frac{\Sigma_{lk} (R^{d*}_{l2}M^{RL}_{d_{lk}}
N_{lk} L^{d}_{k1})^*}{f}\Bigg)^2 \,,\label{kkf} \eea where
$M^{RL}_{d_{ji}}$ is the non diagonal quark mass matrices (see the
Appendix) and $N_{ji}$ is a multiplicity factor.

A comparison  between \eq{mslexp} and \eq{msl} with \eq{kkopfl}
indicates
    that we need $f$ at least\be
f\simeq 10 \,\mbox{TeV},\ee in order to satisfy the experimental
bound.

 However, as
we shall see in the next sections, processes mediated by flavons
are not the dominant ones and the limit obtained here must be
increase. For this reason we will not further discuss this kind of
processes, and concentrate on those mediated by the flavor gauge
bosons.
\section{Processes mediated by the gauge flavons}

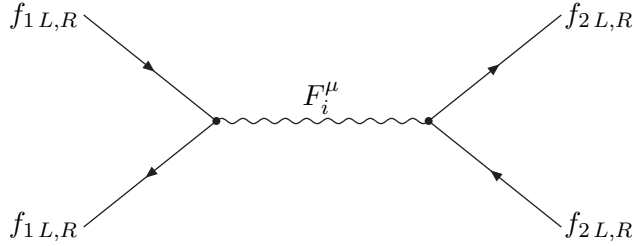
\begin{figure}
\begin{picture}(300,110)(0,0)\Vertex(100,60){1.5}\Vertex(180,60){1.5} \Photon(100,60)(180,60){1}{10}
\Text(140,63)[b]{$F_i^\mu$}
\ArrowLine(50,100)(100,60)\Text(48,105)[r t]{$f_{1\,L,R}$}
\ArrowLine(180,60)(230,100) \Text(232,105)[l t]{$f_{2\,L,R}$}
\ArrowLine(100,60)(50,20)\Text(48,15)[r b]{$f_{1\,L,R}$}
\ArrowLine(230,20)(180,60)\Text(232,15)[l b]{$f_{2\,L,R}$}
\end{picture}
\caption{Processes of annihilation and production of $f\bar{f}$
mediated by gauge flavons.} \label{anni}\end{figure}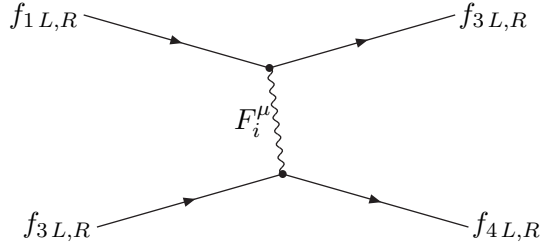
\begin{figure}
\begin{picture}(300,110)(0,0)\Vertex(120,80){1.5}\Vertex(125,40){1.5} \Photon(120,80)(125,40){1}{8}
\Text(107,60)[l]{$F_i^\mu$}
\ArrowLine(50,100)(120,80)\Text(48,105)[r t]{$f_{1\,L,R}$}
\ArrowLine(120,80)(190,100)\Text(192,105)[l t]{$f_{3\,L,R}$}
\ArrowLine(55,20)(125,40) \Text(53,15)[r b]{$f_{3\,L,R}$}
\ArrowLine(125,40)(195,20) \Text(197,15)[l b]{$f_{4\,L,R}$}
\end{picture}
\caption{Parity violation processes mediated by gauge
flavons.}\label{fcnc}\end{figure}

Processes mediated by the gauge bosons of the flavor groups are
crucial in fixing  the scale $f$.
Most of the processes we  discuss in the following arise from two classes of operators of the general form
\bea
\frac{1}{\tilde{\Lambda}^2}\big(\bar{f_{1}}\Gamma^{V,A,S,P}f_1\big)\big(\bar{f_{2}}\Gamma_{V,A,S,P}f_2\big)& \mbox{and}&  \frac{1}{\tilde{\Lambda}^2}\big(\bar{f_{1}}\Gamma^{V,A,S,P}f_2\big)\big(\bar{f_{3}}\Gamma_{V,A,S,P}f_{4}\big)\,,
\label{op}
\eea
where in the second class of operators at least $f_2\neq f_1$ (or  $f_4\neq f_3$ ). These operators arise from integrating out the gauge flavons. Notice that the longitudinal components of the gauge flavons propagators give contributions sub-leading with respect to that arising from the transverse components. The first class of operators in \eq{op} gives rise to processes of annihilation and production of fermion-antifermion couples and  parity violation processes (see \fig{anni}), while the latter to flavor changing processes (see \fig{fcnc}). Tree level processes can give only the vectorial and the axial structure, while the scalar and pseudoscalar ones arise when we consider processes at least at one-loop. For this reason these structures are suppressed and we neglect them in the following.

\subsection{$f\bar{f}\rightarrow f'\bar{f'}$ and parity violation}

The number of four fermions operators belonging to the first class of \eq{op} which give rise to $f\bar{f}\rightarrow f'\bar{f'}$ and parity violation is very large, so we consider only those that contribute to the experimentally most constrained processes, that is, $e^+e^-\rightarrow e^+e^-$ and $e\, q_{L,R}\rightarrow e\, q_{R,L}$. They can be parametrized as
\bea
&&-\frac{1}{f^2}\Big\{\eta_{LL}^{ee}\Big(\bar{e}_{L}\gamma_{\mu}e_{L}\,\bar{e}_{L}\gamma^{\mu}e_{L}\Big)+
\eta_{RR}^{ee}\Big(\bar{e}_{R}\gamma_{\mu}e_{R}\,\bar{e}_{R}\gamma^{\mu}e_{R}\Big)+
 2\eta_{LR}^{ee}\Big(\bar{e}_{L}\gamma_{\mu}e_{L} \,\bar{e}_R\gamma^{\mu}e_{R}\Big)+ \nn\\
&&\eta_{L}^e\eta_L^u \Big(\bar{e}_{L}\gamma_{\mu}e_{L}\,\bar{u}_{L}\gamma^{\mu}u_{L}\Big) +\eta_{R}^e\eta_R^u \Big(\bar{e}_{R}\gamma_{\mu}e_{R}\,\bar{u}_{R}\gamma^{\mu}u_{R}\Big)+ \eta_{L}^e\eta_R^u \Big(\bar{e}_{L}\gamma_{\mu}e_{L}\,\bar{u}_{R}\gamma^{\mu}u_{R}\Big) +
\nn \\
&& \eta_{R}^e\eta_L^u \Big(\bar{e}_{R}\gamma_{\mu}e_{R}\,\bar{u}_{L}\gamma^{\mu}u_{L}\Big)  + (u\rightarrow d)\Big]\Big\} \,.
\label{contactandparity}
\eea
 The first line of \eq{contactandparity} has to be compared with the usual effective lagrangian of contact interactions \cite{parita}
\be
\frac{g^2}{\Lambda^2_{LL}}\Big(\pm\bar{e}_{L}\gamma_{\mu}e_{L}\,\bar{e}_{L}\gamma^{\mu}e_{L}\Big)+
\frac{g^2}{\Lambda^2_{RR}}\Big(\pm\bar{e}_{R}\gamma_{\mu}e_{R}\,\bar{e}_{R}\gamma^{\mu}e_{R}\Big)+
 \frac{g^2}{\Lambda^2_{LR}}\Big(\pm\bar{e}_{L}\gamma_{\mu}e_{L} \,\bar{e}_R\gamma^{\mu}e_{R}\Big)+ \frac{g^2}{\Lambda^2_{RL}}\Big(\pm\bar{e}_{R}\gamma_{\mu}e_{R} \,\bar{e}_L\gamma^{\mu}e_{L}\Big)\nn
\ee
where the limits on $\Lambda_{UU}$, with $U= L,R$, are usually given imposing $g^2= 4\pi$. From \cite{contact} we have
\bea
\Lambda^+_{LL}=8.3\, \mbox{TeV} &\mbox{and}&  \Lambda^-_{LL}=10.1\, \mbox{TeV}\,.
\eea
To compare these values with \eq{contactandparity}, we write the $\eta$ coefficients in terms of the model parameters as
\bea
\eta^{ee}_{LL} &=& \frac{(y^{e11}_{1_{L}})^2}{\epsilon_1^2} + \frac{(y^{e11}_{2_{L}})^2}{\epsilon_2^2}+2\frac{(y^{e11}_{3_{L}})^2}{\epsilon_1^2 + \epsilon_2^2} \,,\nn\\
\eta^{ee}_{RR} &=& \frac{(y^{e11}_{1_{R}})^2}{\epsilon_1^2} + \frac{(y^{e11}_{2_{R}})^2}{\epsilon_2^2}+2\frac{(y^{e11}_{3_{R}})^2}{\epsilon_1^2 +\epsilon_2^2} \,,\nn \\
\eta^{ee}_{LR} &=&
\frac{y^{e11}_{1_{L}}y^{e11}_{1_{R}}}{\epsilon_1^2} +
\frac{y^{e11}_{2_{L}}y^{e11}_{2_{R}}}{\epsilon_2^2}+2\frac{y^{e11}_{3_{L}}y^{e11}_{3_{R}}}{\epsilon_1^2
+ \epsilon_2^2},\nn \eea where $y^{e\alpha \beta}_{m_{U}}$
have been defined in \eq{yn}. A direct comparison imposes \be
f\geq36 \,\mbox{TeV}, \ee which is two order of magnitude bigger
than the  value we found in the previous section for LFV.

Let us turn now to parity violation processes. Parity violation is measured in term of the weak charge $Q_W$ and the most recent experimental values give \cite{parita}
\be
\Delta Q_{w}=0.44\pm0.44 \,.
\label{deltaQv}
\ee
From the contact parameters, $\Delta Q_{w}$ receives the contributions \cite{parita}
 \be
\Delta Q_{w}= (-11.4 \mbox{TeV}^2)(-\tilde{\eta}^{eu}_{LL}+\tilde{\eta}^{eu}_{RR}-\tilde{\eta}^{eu}_{LR}+\tilde{\eta}^{eu}_{RL}) +  (-12.8 \mbox{TeV}^2)(-\tilde{\eta}^{ed}_{LL}+\tilde{\eta}^{ed}_{RR}-\tilde{\eta}^{ed}_{LR}+\tilde{\eta}^{ed}_{RL}) \,,
\label{deltaQe}
\ee
where
\be
\tilde{\eta}^{eq}_{AB}=\frac{4\pi}{\Lambda^{2\,eq}_{AB}}\eta^{e}_{A}\eta^q_B \,.
\ee
In \eq{contactandparity} $4\pi/\Lambda^{2\,eq}_{AB}=-1/f^2$ and the $\eta$ coefficients are given by
\bea
\eta^{e}_L &=& \Bigg(\frac{y^{e11}_{1_{L}}}{\epsilon_1^2} + \frac{y^{e11}_{2_{L}}}{\epsilon_2^2} \Bigg)\,,\nn\\
\eta^{e}_R &=& \Bigg(\frac{y^{e11}_{1_{R}}}{\epsilon_1^2} + \frac{y^{e11}_{2_{R}}}{\epsilon_2^2} \Bigg)\,,\nn \\
\eta^{u}_L &=& y^{u11}_L \,,\nn\\
\eta^{u}_R &=& y^{u11}_R \,,\nn \\
\eta^{d}_L &=& y^{d11}_L\,,\nn\\
\eta^{d}_R &=&  y^{d11}_R\,,\nn \eea where
$y^{e\alpha \beta}_{m_{U}}$ and $y^{q\alpha \beta}_U$ are
given in \eqs{yn}{ynq}. A direct comparison of \eq{deltaQv} with
\eq{deltaQe} gives \be
 f\geq 88 \, \mbox{TeV}.
\label{par}
 \ee
\subsection{Leptonic processes}
The most stringent experimental limits for LFV processes comes from the processes $\mu\rightarrow 3 e$ and $\mu\rightarrow e \gamma$, but in the little-flavon model only $\mu\rightarrow 3e$ is present at tree level. As already discussed in \sec{fl} the limit on the branching ratio for muon decay LFV is given by $\Gamma_{\mu\rightarrow 3e}/\Gamma_{\mu\rightarrow \mbox{\small all}}< 10^{-12}$.

 As done in \eq{parmu} we parametrize the effective
interactions as \be -\frac{1}{
f^2}\Big(g_{LL}\big(\bar{e}_{L}\gamma_{\mu}
\mu_{L}\,\bar{e}_{L}\gamma^{\mu} e_{L}\big)+
g_{RR}\big(\bar{e}_{R}\gamma_{\mu}
\mu_{R}\,\bar{e}_{R}\gamma^{\mu} e_{R}\big)
+g_{LR}\big(\bar{e}_{L}\gamma_{\mu}
\mu_{L}\,\bar{e}_{R}\gamma^{\mu} e_{R}\big)+
g_{RL}\big(\bar{e}_{R}\gamma_{\mu}
\mu_{R}\,\bar{e}_{L}\gamma^{\mu} e_{L}\big) \Big) \,, \ee where
 \bea
g_{LL}&=& \frac{y^{e12}_{1_{L}}y^{e11}_{1_{L}}}{\epsilon_1^2}
+\frac{y^{e12}_{2_{L}}y^{e11}_{2_{L}}}{\epsilon_2^2}+
2\frac{y^{e12}_{3_{L}}y^{e11}_{3_{L}}}{\epsilon_1^2 +
\epsilon_2^2}\,,\nn \\
g_{RR}&=& \frac{y^{e12}_{1_{R}}y^{e11}_{1_{R}}}{\epsilon_1^2}
+\frac{y^{e12}_{2_{R}}y^{e11}_{2_{R}}}{\epsilon_2^2}+2\frac{y^{e12}_{3_{R}}y^{e11}_{3_{R}}}{\epsilon_1^2
+ \epsilon_2^2}\,,\nn \\
g_{LR}&=& \frac{y^{e12}_{1_{L}}y^{e11}_{1_{R}}}{\epsilon_1^2}
+\frac{y^{e12}_{2_{L}}y^{e11}_{2_{R}}}{\epsilon_2^2}+2\frac{y^{e12}_{3_{L}}y^{e11}_{3_{R}}}{\epsilon_1^2+\epsilon_2^2}\,,\nn \\
g_{RL}&=& \frac{y^{e12}_{1_{R}}y^{e11}_{1_{L}}}{\epsilon_1^2}
+\frac{y^{e12}_{2_{R}}y^{e11}_{2_{L}}}{\epsilon_2^2}+2\frac{y^{e12}_{3_{R}}y^{e11}_{3_{L}}}{\epsilon_1^2+\epsilon_2^2}\,.\nn 
\eea 
The rate decay for this process is then given by
 \be
\Gamma^{ gauge}_{\mu\rightarrow 3 e }= \frac{ m_\mu^5}{6(16\pi)^3 f^4}\big(|g_{LL}|^2 + |g_{RR}|^2 + |g_{LR}|^2 + |g_{RL}|^2\big)\,
 \ee
and to satisfy the experimental bound we need \be f> 580\,
\mbox{TeV}, \ee which give us the stringent bound so far.

\subsection{$K^0$-$\bar{K}^0$ mixing}
\label{kkmix}

As done in \sec{fl} among all the FCNC processes with $\Delta F=2$
in the quark sector that are responsible of meson-antimeson
oscillations, we will consider only the processes with $\Delta
S=2$ since the best experimental data are related to the splitting
of Kaon masses \cite{deltamls} (see \eq{mslexp}).

Also this time, in order to estimate the contribution of the gauge
flavons to $\Delta m_{LS}$, we have to consider all the possible
effective operators
with $\Delta S = 2$.
To give a more complete analysis, we will take into account also the operators arising at one-loop level. Accordingly we have six main operators that we parametrize as follows
\bea
&&-\frac{1}{\Lambda^2_0}\Big[ \eta_{1}\Big(\bar{d}\gamma_{\mu}(1-\gamma_5)s \,\bar{d}\gamma^{\mu}(1-\gamma_5)s\Big)\,+\, \eta_{2}\Big(\bar{d}\gamma_{\mu}(1+\gamma_5)s \,\bar{d}\gamma^{\mu}(1+\gamma_5)s\Big)+\nn \\
&& + \eta_{3}\Big(\bar{d}\gamma_{\mu}(1-\gamma_5)s \,\bar{d}\gamma^{\mu}(1+\gamma_5)s\Big] -\frac{1}{\Lambda^2_1}\Big[\eta_{4}\Big(\bar{d}(1+\gamma_5)s \,\bar{d}(1-\gamma_5)s\Big)+ \nn \\
&& \eta_{5}\Big(\bar{d}(1-\gamma_5)s \,\bar{d}(1-\gamma_5)s\Big) +\eta_{6}\Big(\bar{d}(1+\gamma_5)s
\,\bar{d}(1+\gamma_5)s\Big)\Big] \,,
\label{kkop}
\eea
where
\bea
\frac{1}{\Lambda_0^2}= \frac{1}{4 f^2}\Bigg(\frac{1}{\epsilon_1^2} + \frac{1}{\epsilon_2^2}\Bigg) &\mbox{and}&
\frac{1}{\Lambda_1^2}\simeq \frac{1}{(4\pi)^2}\frac{m_{b}^2}{4 f^4}\Bigg(\frac{1}{\epsilon_1^2} + \frac{1}{\epsilon_2^2}\Bigg)^2\,,
\eea
and
\bea
\eta_{1}&=& ( y^{d12}_L )^2 \,,\nn \\
\eta_{2}&=& ( y^{d12}_R )^2 \,,\nn \\
\eta_{3}&=& y^{d12}_L y^{d12}_R \,,\nn\\
\eta_{4}&=& \frac{1}{m_b^2}\Big(\Sigma_{ij} R^{d*}_{j1}y_R^{dj} M^{RL}_{d_{ji}} y_L^{di} L^{d}_{i2}\Big)\Big(\Sigma_{nm} R^{d}_{n2}y_R^{dn} M^{RL*}_{d_{nm}} y_L^{dm} L^{d*}_{m1}\Big) \,,\nn \\
\eta_{5}&=& \frac{1}{m_b^2}\Big(\Sigma_{ij} R^{d*}_{j1}y_R^{dj} M^{RL}_{d_{ji}} y_L^{di} L^{d}_{i2}\Big)^2\,,\nn \\
\eta_{6}&=& \frac{1}{m_b^2}\Big(\Sigma_{nm} R^{d}_{n2}y_R^{dn} M^{RL*}_{d_{nm}}
y_L^{dm} L^{d*}_{m1}\Big)^2\,,
\label{kk} 
\eea 
where $y^{d12}_U$ are defined in \eq{ynq}. \eq{kk} gives the relationships between the effective operators of \eq{kkop} and the model parameters and charges. The last three operators
proportional to $\eta_{4,5,6}$ respectively arise from one-loop
box-diagrams  in which gauge flavon bosons are exchanged. These one-loop
  effects  are not the dominant ones since they only require $f$ to be  $\geq 2$ TeV to satisfy the
   experimental limit, as one can check comparing \eq{mslexp} and \eq{msl} with \eq{kk}. The first three operators come  from
    tree level processes and a comparison to the experimental limits indicates that they impose  at least\be
f\simeq 4\times 10^3\,\mbox{TeV}, \ee to satisfy the bound in
\eq{mslexp}. This result shifts the scale we have found in the
lepton sector of more than an order of magnitude and definitively
fixes the lowest scale for the breaking of the global symmetry
that give rise to the little-flavons.

\section{Effects at one loop}

\subsection{Rare processes}
\label{rare}
\begin{figure}
\begin{picture}(300,90)(0,0)\Vertex(90,15){1.5}\Vertex(170,15){1.5} \ArrowArcn(130,15)(40,180,0)
\Text(110,49.6)[c]{$\times$}\Photon(90,15)(170,15){1}{8}\ArrowLine(40,15)(90,15)\Text(38,16)[r
b]{$\mu$} \ArrowLine(170,15)(210,15)\Text(212,16)[l
b]{$e$}\Text(110,55)[r
b]{$e,\mu,\tau$}\Photon(150,49.6)(190,80){1}{6} \Text(170,70)[r
b]{$\gamma$}\Text(130,16)[c b]{$F^\mu_i$}\end{picture}
\begin{picture}(300,90)(0,0)\Vertex(90,15){1.5}\Vertex(170,15){1.5} \ArrowArcn(130,15)(40,180,0)
\Photon(90,15)(170,15){1}{8}\ArrowLine(40,15)(90,15)\Text(70,15)[c]{$\times$}\Text(38,16)[r
b]{$\mu$}\ArrowLine(170,15)(210,15) \Text(212,16)[l b]{$e$}
\Text(110,55)[r b]{$e,\mu,\tau$}\Photon(150,49.6)(190,80){1}{6}
\Text(170,70)[r b]{$\gamma$} \Text(130,16)[c
b]{$F^\mu_i$}\end{picture} \caption{Gauge flavons mediated
contribution to the decay $\mu\rightarrow e\gamma$}. \label{egamma}
\end{figure}
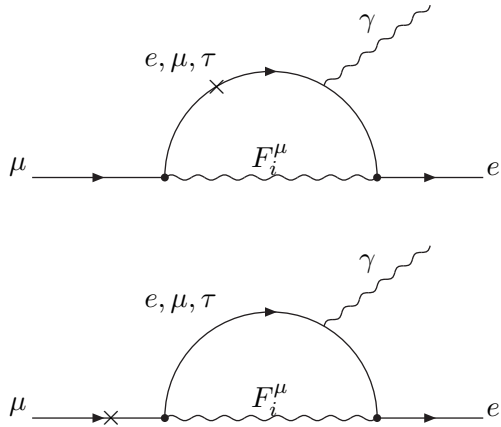
There are some rare decays that in the model occur only at one loop, but give a bound on $f$ which is comparable to the bound obtained from the analysis of the tree-level processes.

 As an example let us consider the LFV process $\mu\rightarrow e\gamma$. For the $\mu\rightarrow e \gamma$ process we have the strong limit\cite{mutree}
\be
 \frac{\Gamma_{\mu\rightarrow e \gamma}}{\Gamma_{\mu\rightarrow \mbox{\small all}}}< 1.2\,\times 10^{-11} \,.
\label{boundegamm}
\ee
 We can parametrize the interaction which gives rise to the decay as
\be \Big(\bar{e}\,i\sigma_{\nu\mu}(1-\gamma_5)\,\mu
\mathcal{M}^{LR}+ \bar{e}\,i\sigma_{\nu\mu}(1+\gamma_5)\,\mu
\mathcal{M}^{RL}\Big)F^{\nu\mu} \,. \label{mueg}
 \ee
In the model we have two kind of diagrams that contribute to the
process $\mu\rightarrow e\gamma$ (see \fig{egamma}). The second  decay in \fig{egamma} is
present also in the Standard Model---with the charged W bosons and massive neutrinos in the loop---and
gives a contribution proportional to $m_{\mu}/m^2_F$, where
$m_F^2$ is the mass of the flavor gauge boson. On the contrary, the
first is not present in the Standard Model and is possible because
the flavor gauge bosons couple also to right handed fermions. It
gives a contribution proportional to
$m_{\alpha}/m^2_F\log(m_{\alpha}^2/m^2_F)$ where $m_\alpha$ is the
mass of the fermion circulating in the loop. For this reason the
dominant contribution comes from the $\tau$ exchange. For this
process we have \bea
 \mathcal{M}^{RL}&=&\frac{e}{2\pi^2}\frac{m_\tau}{f^2}\Bigg(\log\frac{m_\tau^2}{f^2}\Bigg)Y^{RL} \,,\nn\\
 &&\nn\\
\mathcal{M}^{LR}&=&\frac{e}{2\pi^2}\frac{m_\tau}{f^2}\Bigg(\log\frac{m_\tau^2}{f^2}\Bigg)Y^{LR}
\,, \eea where $e$ is the electric charge and \bea Y^{RL}&=&
\frac{y^{e3 2}_{1_{L}}
 y^{e1 3}_{1_{R}}}{\epsilon_1^2}+ \frac{y^{e3 2}_{2_{L}}
 y^{e1 3}_{2_{R}}}{\epsilon_2^2} +2\frac{y^{e32}_{3_{L}}y^{e1 3}_{3_{R}}}
 {\epsilon_1^2+\epsilon_2^2}\, , \nn\\
&& \nn\\
Y^{LR}&=& \frac{y^{e3 2}_{1_{R}}
 y^{e1 3}_{1_{L}}}{\epsilon_1^2}+ \frac{y^{e3 2}_{2_{R}}
 y^{e1 3}_{2_{L}}}{\epsilon_2^2} +2\frac{y^{e32}_{3_{R}}y^{e1 3}_{3_{L}}}
 {\epsilon_1^2+\epsilon_2^2} \,.
\eea
The rate decay for this process is then given by
 \be
\Gamma_{\mu\rightarrow e \gamma}=
\frac{3\alpha}{8\pi^4}\frac{m_\mu^3
m_\tau^2}{f^4}\Bigg(\log\frac{m_\tau^2}{f^2}\Bigg)^2\Big(Y^{RL^{2}}+Y^{LR^{2}}\Big)
 \ee
In order to satisfy the experimental bound of \eq{boundegamm} we
need \be f\simeq4 \times\,10^3\,\mbox{TeV}. \ee which is of the
same order of the value obtained in \sec{kkmix}.

The process corresponding to the LFV process $\mu \rightarrow e
\gamma$ in the quark sector is the FCNC process $b \rightarrow s
\gamma$. For the $b \rightarrow s \gamma$ process we have the
limit \cite{bsg} \be
 \frac{\Gamma_{b\rightarrow s \gamma}}{\Gamma_{b\rightarrow \mbox{\small all}}}< (3.3\pm0.4)\,\times \,10^{-4} \,.
\label{boundsgamm} \ee The Standard Model effective interaction
that is responsible of this process is parametrized as
\cite{buras} 
\be \frac{-4 G_F}{\sqrt{2}}V^{*}_{ts}V_{tb}\frac{e
m_b}{16 \pi^2}C_7(m_W)\bar{s}_L \sigma_{\mu\nu}F^{\mu\nu}b_R\,,
\label{bsgsm}\ee 
where $C_7(m_W)$ is the Wilson coefficient and is
a function of $m_t(m_W)$ and $m_W$ as reported in \cite{buras}. In the following we will neglect of the renormalization effects and we will compare the effective interaction which gives rise to the decay $b\rightarrow s \gamma$ in our model with the one loop electroweak operator of \eq{bsgsm}.
 Analogously to what done for the process $\mu\rightarrow e \gamma$, we  parametrize the interaction responsible of the decay $b \rightarrow s
\gamma$ as \be
\Big(\bar{s}\,i\sigma_{\nu\mu}(1-\gamma_5)\,b \mathcal{N}^{LR}+
\bar{s}\,i\sigma_{\nu\mu}(1+\gamma_5)\,b
\mathcal{N}^{RL}\Big)F^{\nu\mu} \,.
 \label{bsgfl}\ee
All the considerations done for the process $\mu\rightarrow
e\gamma$ may be applied in this context and for this reason the
dominant contribution to the process $b \rightarrow s \gamma$
comes from the loops in which a quark $b$ is exchanged. For the
process we are considering we have \bea
 \mathcal{N}^{RL}&=&\frac{e}{2\pi^2}\frac{m_b}{f^2}\Bigg(\log\frac{m_b^2}{f^2}\Bigg)X^{RL} \,,\nn\\
 &&\nn\\
\mathcal{N}^{LR}&=&\frac{e}{2\pi^2}\frac{m_b}{f^2}\Bigg(\log\frac{m_b^2}{f^2}\Bigg)X^{LR}
\,, \eea where $e$ is the electric charge and \bea X^{RL}&=& y^{d3
3}_{L}
 y^{d2 3}_{R}\Bigg(\frac{1}{\epsilon_1^2}+ \frac{1}{\epsilon_2^2} \Bigg)\, ,\nn\\
X^{LR}&=& y^{d3 3}_{R}
 y^{d2 3}_{L}\Bigg(\frac{1}{\epsilon_1^2}+  \frac{1}{\epsilon_2^2} \Bigg) \,.
\eea A comparison between \eq{bsgsm} and \eq{bsgfl} indicates that
we need \be f\simeq 31\,\mbox{TeV}, \ee in order to have the two
contribution of the same order.

\subsection{Muon anomalous magnetic moment}
\label{gmeno} 
In \sec{rare} we have seen how one loop processes
give a bound on $f$ comparable to that one obtained by tree level
processes. We may ask ourselves what is the limit on $f$ we
obtain if we consider the one loop contribution to the anomalous
magnetic moment of the muon. 
The uncertainty
between the experimental data and the theoretical computation for the anomalous magnetic moment of the muon $a_\mu = (g_\mu-2)/2$ is  \cite{g}
\be \Delta a_\mu= a^{\mbox{\small{exp}}}_\mu-a^{\mbox{\small{SM}}}_\mu=(27\pm 14)\,\times 10^{-10} \,.\label{dis}\ee

In order to obtain a limit for $f$ from \eq{dis} we have to
consider the effective interaction that is proportional to
$a_\mu$. The interaction coincides to that of \eq{mueg} that is
responsible of the rare decay $\mu \rightarrow e \gamma$ once we
substitute the outgoing electron with  an outgoing muon. Therefore
it is parametrized as \be
\Big(\bar{\mu}\,i\sigma_{\nu\mu}(1-\gamma_5)\,\mu
\tilde{\mathcal{M}}^{LR}+
\bar{\mu}\,i\sigma_{\nu\mu}(1+\gamma_5)\,\mu
\tilde{\mathcal{M}}^{RL}\Big)F^{\nu\mu} \,. \label{mug}
 \ee
As in \sec{rare} the dominant contribution comes from $\tau$
exchanging and for this reason we have \bea
 \tilde{\mathcal{M}}^{RL}&=&\frac{e}{2\pi^2}\frac{m_\tau}{f^2}\Bigg(\log\frac{m_\tau^2}{f^2}\Bigg)\tilde{Y}^{RL} \,,\nn\\
 &&\nn\\
\tilde{\mathcal{M}}^{LR}&=&\frac{e}{2\pi^2}\frac{m_\tau}{f^2}\Bigg(\log\frac{m_\tau^2}{f^2}\Bigg)\tilde{Y}^{LR}
\,, \eea where $e$ is the electric charge and \bea \tilde{Y}^{RL}&=&
\frac{y^{e3 2}_{1_{L}}
 y^{e2 3}_{1_{R}}}{\epsilon_1^2}+ \frac{y^{e3 2}_{2_{L}}
 y^{e2 3}_{2_{R}}}{\epsilon_2^2} +2\frac{y^{e32}_{3_{L}}y^{e2 3}_{3_{R}}}
 {\epsilon_1^2+\epsilon_2^2} \, ,\nn\\
&& \nn\\
\tilde{Y}^{LR}&=& \frac{y^{e3 2}_{1_{R}}
 y^{e2 3}_{1_{L}}}{\epsilon_1^2}+ \frac{y^{e3 2}_{2_{R}}
 y^{e2 3}_{2_{L}}}{\epsilon_2^2} +2\frac{y^{e32}_{3_{R}}y^{e2 3}_{3_{L}}}
 {\epsilon_1^2+\epsilon_2^2} \,.
\eea From \eq{mug} we have \be a_\mu=\frac{m_\mu
(\tilde{\mathcal{M}}^{RL}+\tilde{\mathcal{M}}^{LR})}{e}\,.
\label{gg}\ee A direct comparison between \eq{dis} and \eq{gg}
gives \be f\simeq 28\,\mbox{TeV},\ee that does not change the
previous results obtained in \sec{kkmix}.

\section{Stabilization of the electroweak scale}
\label{stab}

\begin{figure}
\begin{picture}(300,90)(0,0)\Vertex(90,40){1.5}\Vertex(150,40){1.5}
\ArrowArcn(120,40)(30,180,0) \ArrowArcn(120,40)(30,0,180)
\DashLine(50,40)(90,40){2}
\DashLine(150,40)(190,40){2}\DashLine(90,40)(70,70){1.5}
\DashLine(90,40)(70,10){1.5} \DashLine(150,40)(170,70){1.5}
\DashLine(150,40)(170,10){1.5} \Text(70,70)[c]{$\times$}
\Text(70,10)[c]{$\times$}\Text(170,70)[c]{$\times$}
\Text(170,10)[c]{$\times$} \Text(68,72)[r]{$\phi_{1,2}$}
\Text(68,8)[r]{$\phi_{1,2}$}\Text(172,72)[l]{$\phi_{1,2}$}
\Text(172,8)[l]{$\phi_{1,2}$} \Text(60,43)[b]{$H$}
\Text(180,43)[b]{$H$} \Text(120,74)[b]{$t_{L}$}
\Text(120,4)[c]{$c_R$}
\end{picture}
\caption{Potentially dangerous fermion one-loop correction to the
Higgs mass.} 
\label{h}
\end{figure}
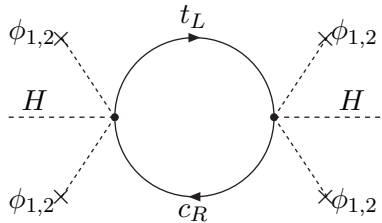

From the little-flavon model point of view the stabilization of the electroweak scale
must be thought as obtained by means of a little-Higgs mechanism acting on the radiative corrections to the
 Higgs mass. However, since the scale of our model turns out to be quite high with respect to that electroweak,  we have to worry
 about the corrections to the Higgs mass coming from the fermionic sector and its interaction
  with the flavons. The potentially dangerous contributions come from one and more loop diagrams which are quadratic
  divergent and in which the cut-off should be taken at $\Lambda\simeq 5\times 10^4$ TeV. Usually
   little-Higgs models protect the Higgs mass from the quadratic divergences arising from one-loop
   corrections up to a scale $\Lambda_H\simeq 10$ TeV \cite{littlehiggs} and
   $f_H\simeq \Lambda_H/4\pi\simeq 1$ TeV is the scale at which the Higgs arises
   as a pseudo-Goldstone boson. It is possible to protect the Higgs mass even from the quadratic corrections arising from fermions two-loops if the approximate global symmetries of the little-Higgs model are enlarged. The cut-off of such a model is then shifted at $\Lambda_H\simeq 100$ TeV. In \cite{Nelson} an example of how this mechanism works is given.

The scale of the little-flavon model is so large that it forces us  to protect the Higgs mass up to four-loops quadratic divergences and this is done by further enlarging the global symmetries of the little-Higgs model. Since the choice of the little-Higgs model is independent of the little-flavon model, we shall give an example of how this mechanism is applied to the Littlest Higgs model that is based on an $SU(5)$ global symmetry ( first reference in Ref.\cite{littlehiggs}).

For each Standard Model family, both for quarks and the leptons,  $U_i$ and $U^c_i$, we add four quintuplet, $X_{i},\, X^c_i,\, Y_i,\, Y^c_i$, in order to have an approximate global symmetry $\big[SU(5)\big]^5$. Only $X^c_i$ and $X_i$ are charged under the flavor gauge group $SU(2)_F\times U(1)_F$ and their charges are that of the corresponding Standard Model fermions $U_i$ and $U^c_i$.  The Yukawa lagrangian is then given only by terms that leave invariant at least one of the five  $SU(5)$ and takes the following expression
\be
\mathcal{L}_Y = \lambda_1 f_H A_{ij} X_i\Sigma_H X_j^c + \lambda_2 f_H X_j^c Y_j + \lambda_3 f_H Y_j Y^c_j + \lambda_4 f_H Y^c_j U_j +\lambda_5 f_H X_i U^c_i \,,
\label{new}
\ee
where with $A_{ij}$ we indicate the flavons which couple to $X_i$ and $X_j^c$ and with $\Sigma_H$ the non linear representations of the little-Higgs model pseudo-Goldstone bosons. In this way the Higgs mass receives a quadratic divergence only from the diagrams in which all the $SU(5)$ approximate global symmetries are broken and this happens only from five loops on. The light fermions, $\tilde{q}_j$ and $\tilde{u}^c_i$, are now given by appropriate combinations of the fields appearing in \eq{new} and the effective coupling between light fermions, Higgs and flavons is then given by
\be
\mathcal{L}_Y =\frac{\lambda_1\lambda_2\lambda_3\lambda_4\lambda_5}{\sqrt{\lambda_1^2+\lambda_2^2}\sqrt{\lambda_1^2+\lambda_5^2}\sqrt{\lambda_3^2+\lambda_4^2}\sqrt{\lambda_2^2+\lambda_3^2}}A_{ij}\tilde{u}^c_i \tilde{q}_j h \,.
\label{newc}
\ee
The expression for $\mathcal{L}_Y$ reported in the Appendix must therefore be read as effective terms, the explicit form of which is like that in \eq{newc}.

\section{Conclusion}
While the results  obtained in \cite{noi} are independent of the scale $f$ and the masses and mixing matrices are determined only by the value of $\epsilon_{1}$ and $\epsilon_2$, a detailed analysis of the flavor changing processes induced by the new particles introduced by the little-flavon model shows that this model has to live at least at a scale $\Lambda = 4\pi f \simeq 5\times 10^4$ TeV.

This result has two consequences for the little-flavon model. On
the one hand, the determination of a bound on the scale $f$ leads
to a specific prediction for the scale for the see-saw mechanism
that in the model is used to give mass to the neutrinos. The value
found of $10^4$ TeV  allows to have the couplings of the Dirac
neutrino mass term to be of the same order of the charged lepton
ones. As consequence these couplings are of order $10^{-2}$, two
order of magnitude below those of the previous rough estimate
\cite{noi}.

On the other hand, the scale of the model turns out to be quite
high with respect to that of the electroweak symmetry breaking,
thus making rather challenging a hypothetical unification of the
little-flavon and the little-Higgs models in a single gauge  and
flavor symmetry scenario. However, the high scale of the little
flavon model is not dangerous for the electroweak scale since it
is possible to keep the latter stable ---in the fermion sector---
up to the scale $\Lambda \simeq 5\times 10^4$ TeV, as argued in \sec{stab}.

\section{Acknowledgment}
It is a pleasure to thanks S.\ Bertolini, M.\ Fabbrichesi and M.\ Piai  for help and
very useful discussions, and the CERN Theoretical Division for the hospitality during the completion of this work. This work is
partially supported by  the European TMR Networks HPRN-CT-2000-00148
and HPRN-CT-2000-00152.


\newpage

\section*{APPENDIX}
The Yukawa lagrangian $\mathcal{L}_Y$ is  rather cumbersome. Here we give its expression as a function of $\Sigma=\exp(i\Pi/f)\Sigma_0$ and the charged lepton and quark fields as discussed in \cite{noi}. In this context we shall neglect the neutrino term since neutrinos do not enter in any process we have considered. We also give the charge assignments for all the fields that enter in $\mathcal{L}_Y$ (see \tab{charge}).
\be
\mathcal{L}_{Y}= \mathcal{L}_{u}+\mathcal{L}_{d}+\mathcal{L}_{e} \,,
\ee
where
\bea
-\mathcal{L}_{u} &=&
\lambda_{31}\overline{t_{R}}\big(\Sigma_{\alpha-1\,3}\Sigma_{3\,2+\alpha}\big)^3\tilde{H}^\dag Q_{1L} +\lambda_{32}\overline{t_{R}}\big(\Sigma_{\alpha-1\,3}\Sigma_{3\,2+\alpha}\big)^2\tilde{H}^\dag Q_{2L} \nn \\
&+&\overline{t_{R}}(\lambda_{33} + \lambda_{33}'\Sigma_{\alpha-1\,6}\Sigma_{3\,2+\alpha}+ \lambda_{33}''\Sigma_{\alpha-1\,3}\Sigma_{6\,2+\alpha}\big)\tilde{H}^\dag Q_{3L}\nn \\
&+ &
\lambda_{21}\overline{c_{R}}\big(\Sigma_{\alpha-1\,3}\Sigma_{3\,2+\alpha}\big)^4\tilde{H}^\dag Q_{1L} +\lambda_{22}\overline{c_{R}}\big(\Sigma_{\alpha-1\,3}\Sigma_{3\,2+\alpha}\big)^3 \tilde{H}^\dag Q_{2L} \nn \\
&+& \lambda_{23}\overline{c_{R}}\big(\Sigma_{\alpha-1\,3}\Sigma_{3\,2+\alpha}\big) \tilde{H}^\dag Q_{3L}\nn \\
&+ &
\lambda_{11}\overline{u_{R}}\big(\Sigma_{\alpha-1\,3}\Sigma_{3\,2+\alpha}\big)^6\tilde{H}^\dag Q_{1L} +\lambda_{12}\overline{u_{R}}\big(\Sigma_{\alpha-1\,3}\Sigma_{3\,2+\alpha}\big)^5\tilde{H}^\dag Q_{2L}\nn \\
&+& \lambda_{13}\overline{u_{R}}\big(\Sigma_{\alpha-1\,3}\Sigma_{3\,2+\alpha}\big)^3 \tilde{H}^\dag Q_{3L}\, +\, H.c. \nn\\
&&\nn\\
-\mathcal{L}_{d} &=&
\tilde{\lambda}_{31}\overline{b_{R}}\big(\Sigma_{\alpha-1\,3}\Sigma_{3\,2+\alpha}\big)^5 H^\dag Q_{1L} +\tilde{\lambda}_{32}\overline{b_{R}}\big(\Sigma_{\alpha-1\,3}\Sigma_{3\,2+\alpha}\big)^4 H^\dag Q_{2L} \nn \\
&+ &
\tilde{\lambda}_{33}\overline{b_{R}}\big(\Sigma_{\alpha-1\,3}\Sigma_{3\,2+\alpha}\big)^2 H^\dag Q_{3L}\nn \\
&+ &
\tilde{\lambda}_{21}\overline{s_{R}}\big(\Sigma_{\alpha-1\,3}\Sigma_{3\,2+\alpha}\big)^5 H^\dag Q_{1L} +\tilde{\lambda}_{22}\overline{s_{R}}\big(\Sigma_{\alpha-1\,3}\Sigma_{3\,2+\alpha}\big)^4 H^\dag Q_{2L} \nn \\
&+& \tilde{\lambda}_{23}\overline{s_{R}}\big(\Sigma_{\alpha-1\,3}\Sigma_{3\,2+\alpha}\big)^2H^\dag Q_{3L}\nn \\
&+ &
\tilde{\lambda}_{11}\overline{d_{R}}\big(\Sigma_{\alpha-1\,3}\Sigma_{3\,2+\alpha}\big)^7 H^\dag Q_{1L}
+\tilde{\lambda}_{12}\overline{d_{R}}\big(\Sigma_{\alpha-1\,3}\Sigma_{3\,2+\alpha}\big)^6 H^\dag Q_{2L}\nn \\
&+& \tilde{\lambda}_{13}\overline{d_{R}}\big(\Sigma_{\alpha-1\,3}\Sigma_{3\,2+\alpha}\big)^4 H^\dag Q_{3L}\, +\, H.c.\nn\\
&&\nn\\
\mathcal{L}_{e}
&=& \overline{e_{R}}\ \left[ \lambda_{1e}\
(\Sigma_{\alpha-1\,6}\Sigma_{6\,2+\alpha})^{(-Y_{1L}+Y_{1R})}\
(H^\dag\ {l}_{1L})
\right.  \nn \\
&& \quad \quad
+ \left.
i\ (\lambda_{3e}\  \Sigma_{6\,2+\alpha} + \lambda_{2e}\  \epsilon_{\alpha\,\beta} \Sigma_{\beta-1\,6})(\Sigma_{\delta-1\,6}\Sigma_{6\,2+\delta})^{(Y_{1R}-1)}\
(H^\dag\ {l}_{\alpha L}) \right]  \nn \\
&+&  \overline{E_{\alpha R}}\Big[ i
\left( \lambda_{1E}'\  \Sigma_{6\,2+\alpha} + \lambda_{1E}\ \epsilon_{\alpha\,\beta} \Sigma_{\beta -1\,6} \right)
(\Sigma_{\delta-1\,6}\Sigma_{6\,2+\delta})^{-Y_{1L}} \Big]
(H^\dag\ {l}_{1L})  \nn \\
&+&  \overline{E_{\alpha R}}\Big[
\delta_{\alpha\,\beta} \big(
- \lambda_{2E}\ + \lambda_{2E}'\ \Sigma_{\gamma-1\,6}\Sigma_{3\,2+\gamma}
+ \lambda_{2E}''\  \Sigma_{\gamma-1\,3}\Sigma_{6\,2+\gamma} \big)
  \nn \\
&& 
+ \Big(\lambda_{3E}\  \Sigma_{\alpha-1\,6}\Sigma_{3\,2+\beta}
+ \lambda_{3E}' \ \epsilon_{\alpha\,\delta}\epsilon_{\beta\,\gamma}\ \Sigma_{6\,2+\delta}\Sigma_{\gamma-1\,3} + (3\leftrightarrow 6) \Big) \nn \\
&& 
+\Big( \lambda_{4E}\  \Sigma_{\alpha-1\,6}\Sigma_{\gamma-1\,3}
\epsilon_{\beta\,\gamma}
+ \lambda_{4E}'\ \epsilon_{\alpha\,\delta}
\Sigma_{6, 2+\delta}\Sigma_{3, 2+\alpha}+ (3\leftrightarrow 6)
\Big)
\Big] (H^\dag\ {l}_{\beta L})\  + H.c.\,.
\label{yuk}
\eea
\vfill
\begin{table}[ht]
\begin{center}
\caption{Charges of fermion  and flavon fields ($\alpha= 2,3$)
under the horizontal flavor groups $SU(2)_F$ and $U(1)_F$.
${l}_{f L}$ and $Q_{iL}$ stands for the electrweak
left-handed doublets. $q$ is an arbitrary not determined charge and being universal in the quark sector does not affect any process.}
\label{charge}
\vspace{0.2cm}
\begin{tabular}{|c|c|c|}
\hline
&\quad\quad $U(1)_F$ \quad\quad &\quad $SU(2)_F$ \quad \cr
\hline
${l}_{e L}$ &   $-2$ & 1 \cr
$e_R$  & $2$ & 1 \cr
${L}_L = ( {l_{\mu}}\, ,\,{l_{\tau}})_L$  & $1/2$ & 2  \cr
$E_R = (\mu \,,\,\tau)_R$  & $1/2$ & 2  \cr
$\nu_{1R}$  & $1$ & 1  \cr
$\nu_{2R}$  & $-1$ & 1  \cr
$\nu_{3R}$  & $0$ & 1  \\
\hline
$Q_{1L}$ &   $q+3$ & 1 \cr
$Q_{2L}$  & $q+2$ & 1 \cr
$Q_{3L}$  & $q$ & 1 \cr
$u_{R}$  & $q-3$ & 1  \cr
$c_{R}$  & $q-1$ & 1  \cr
$t_{R}$  & $q$ & 1  \cr
$d_{R}$  & $q-4$ & 1  \cr
$s_{R}$  & $q-2$ & 1  \cr
$b_{R}$  & $q-2$ & 1  \\
\hline
$\Sigma_{\alpha-1\, 6} = (- i/f\ \phi_1 + ...)_{\alpha-1}$ &1/2 & 2 \\
$\Sigma_{\alpha-1\, 3} = (+ i/f\ \phi_2 + ...)_{\alpha-1}$ &$-1/2$ & 2 \\
$\Sigma_{3\, 2+\alpha} = (- i/f\ \phi_1^* + ...)_{\alpha-1}$ & $-1/2$&  $2^*$ \\$\Sigma_{6\, 2+\alpha} = (- i/f\ \phi_2^* + ...)_{\alpha-1}$ &1/2 &  $2^*$ \\
\hline
\end{tabular}
\end{center}
\end{table}
The charged lepton and quark mass matrices in the current basis are given by
\begin{widetext}
 \be
{ M}^{RL}_e =   \langle h_0 \rangle
\left( \begin{array}{c c c}  \lambda_{1e}\, \varepsilon_{1}^4\varepsilon_{2}^4 & \lambda_{2e}\, \varepsilon_{1}^2 \varepsilon_2 & \lambda_{3e} \, \varepsilon_1 \varepsilon_{2}^2  \\
 \lambda_{1E} \,\varepsilon_{1}^{2} \varepsilon_{2}^3 & \lambda_{2E}  &
 (\lambda_{14E}^\prime + \lambda_{24E}^\prime) \,\varepsilon_{1}\varepsilon_{2}   \\
 \lambda_{1E}' \,\varepsilon_{1}^{3} \varepsilon_{2}^{2} &
 - (\lambda_{14E} + \lambda_{24E}) \,\varepsilon_{1}\varepsilon_{2}  &
 \lambda_{2E}   \\
\end{array} \right)
\label{charged-mass} \, ,
\ee
\end{widetext}
where the notation follows that of eq.~(39) in ref.~\cite{noi}, and
\bea
{M}^{RL}_{u} = \langle h_0  \rangle
\left( \begin{array}{c c c}
 \lambda_{11} k^6 &  \lambda_{12} k^5& \lambda_{13}  k^3 \\
 \lambda_{21} k^4 &  \lambda_{22} k^3  & \lambda_{23} k   \\
 \lambda_{31} k^3 &  \lambda_{32} k^2 &  \lambda_{33}
\end{array} \right)
\label{mass-up0}
& \mbox{and} &
{M}^{RL}_{d} = \langle h_0  \rangle \, k^2 \,
\left( \begin{array}{c c c}
\tilde \lambda_{11} k^5 &  \tilde \lambda_{12} k^4 & \tilde  \lambda_{13}k^2 \\
\tilde  \lambda_{21}k^3 & \tilde  \lambda_{22} k^2  & \tilde \lambda_{23}   \\
\tilde  \lambda_{31}k^3 & \tilde  \lambda_{32}k^2  & \tilde  \lambda_{33}
\end{array} \right) \, ,
\label{mass-down0}
\eea
where $k = \epsilon_1 \epsilon_2$ .
\end{document}